\documentstyle[12pt]{report}
\begin{document}
\begin{centering}
\large\bf A complementary group technique for a \\
resolution of the outer multiplicity problem of $SU(n)$:\\
(I) Littlewood rule and a complementary group of $SU(n)$
\vskip .6truecm
\normalsize Feng Pan$^{\dagger}$ and J. P. Draayer
\vskip .2cm
\noindent{\small\it Department of Physics {\normalsize\&} Astronomy, 
Louisiana State University,}\\
{\small\it Baton Rouge, LA 70803-4001}\\
\vskip 1truecm
{\bf Abstract}\\
\end{centering}
\vskip .5cm
\normalsize A complementary group to $SU(n)$ is found that realizes all features of the Littlewood
rule for Kronecker products of $SU(n)$ representations. This is  accomplished by considering a state
of $SU(n)$ to be a special Gel'fand state of  the complementary group ${\cal U}(2n-2)$. The labels of
${\cal U}(2n-2)$ can be  used as the outer multiplicity labels needed to distinguish multiple
occurrences  of irreducible representations (irreps) in the $SU(n)\times SU(n)\downarrow SU(n)$  
decomposition that is obtained from the Littlewood rule. 
Furthermore, this realization can be used to determine
$SU(n)\supset SU(n-1)\times  U(1)$ Reduced Wigner Coefficients (RWCs) and Clebsch-Gordan  Coefficients
(CGCs) of
$SU(n)$, using algebraic or numeric methods, in either the canonical or a noncanonical basis. The
method is recursive in that it uses simpler  RWCs or CGCs with one symmetric  irrep in conjunction
with standard recoupling procedures. New explicit formulae for the multiplicity  for $SU(3)$ and
$SU(4)$ are used to illustrate the theory.
\vskip 4.5cm
\noindent {-----------------------------------------}\\
\noindent $^{\dagger}$On leave from Department of Physics,
Liaoning Normal Univ., Dalian 116029, P. R.~~China

\newpage

\begin{centering}
{\large I. Introduction}\\
\end{centering}
\vskip .4truecm
The Reduced Wigner Coefficients (RWCs) of $SU(n)\supset SU(n-1)
\times U(1)$ are of importance in many physical applications. Except for those of $SU(2)$, which have
been discussed extensively and expressed in various forms, RWCs of $SU(n)\supset SU(n-1)\times U(1)$,
which can be used to evaluate CGCs of SU(n) in its canonical basis according to the Racah
factorization lemma, have only been given analytically for some special cases. The biggest challenge
involves the outer multiplicity in the decomposition of Kronecker products of $SU(n)\times
SU(n)\downarrow SU(n)$. The first non-trivial but simplest $n=3$ case was studied as part of the
first applications of  non-multiplicity-free CGCs of $SU(3)$ in nuclear and particle physics. There
are several very distinct approaches to the problem: (i) a tensor operator method; (ii) an
infinitesimal generator approach, in which matrix elements of $SU(n)$ generators are used to
determine recursion relations for the RWCs and CGCs; (iii) a polynomial basis and generating
invariants, in which a convenient model space is used to realize the basis of irreps; and (iv) use of
the Schur-Weyl duality relation between $SU(n)$ and the symmetric group $S_{f}$.  Among these are
several ways of solving the problem; indeed, sometimes a combination of two or more methods is used.
There are also different schemes for handling the outer multiplicity, especially for $SU(3$), and
these are usually referred to as either the canonical or a noncanonical labeling scheme.
\vskip .3cm
   A very thoroughly discussed approach to this problem is the canonical unit tensor operator method
developed by Biedenharn and collaborators in a series of publications.$^{[1-8]}$  The unit tensor
operator approach is particularly useful for deriving multiplicity-free CGCs of $U(n)$. The techniques
that are part of this method have also proven to be useful in other approaches, but the method has not
been used to produce a closed algebraic solution to the general outer multiplicity problem. This method
was revisited in the late eighties in a Bargmann Hilbert space representation using the Vector Coherent
State (VCS) theory.$^{[9-11]}$ Although the results seem no simpler than those found earlier, they do
show that there is a relationship between $U(3)\supset U(2)$ RWCs and 3nj coefficients of $SU(2)$,
with some of these being a consequence of the Schur-Weyl duality relation between the unitary and
symmetric groups given by Ali\v{s}uaskas et al.$^{[12-14]}$ 
\vskip .3cm
   Noncanonical definitions of $SU(n)$ outer multiplicity labels, especially of $SU(3)$, have also been
discussed rather extensively, for example by Moshinsky et al,$^{[15-16]}$ Derome and Sharp,$^{[17-18]}$
Resnikoff,$^{[19]}$ Pluha\v{r} et al.$^{[20-21]}$ A wider class of RWCs has been considered by
Hecht,$^{[22]}$ Klimyk and Gavrilik,$^{[23]}$ and Le Blanc and Rowe,$^{[24]}$ who used definitions
related to the canonical scheme. Generally, however, these results are for noncanonical labeling schemes. A
further example is the extensive work of Ali\v{s}auskas$^{[25-28]}$, who investigated paracanonical
coupling relations and symmetries and various pseudo-canonical coupling schemes, which lead to
biorthogonalities among the corresponding coefficients. It should be stated that noncanonical
definitions for $SU(n)$ coupling coefficients normally lead to non-orthogonality with respect to the
outer multiplicity. In such cases, the Gram-Schmidt process can be adopted to recover orthonormality,
but this procedure includes an arbitrary choice in ordering the elements to be orthogonalized.
Generally, only a  numerical algorithm is possible except a few simple cases where analytical
expressions are available.$^{[24-28]}$
\vskip .3cm
   The Schur-Weyl duality relation between $SU(n)$ and $S_{f}$ was also used by several authors. It was
first studied by Moshinsky,$^{[16]}$ Kramer,$^{[29]}$ and Alisauskas and Jucy,$^{[14-16]}$ who were 
able to demonstrate that the scheme works in the multiplicity-free and non-multiplicity-free cases. For
non-multiplicity-free couplings, however, numerical orthogonalizition is required. This is illustrated 
for some simple cases in the work of Chen et al,$^{[30]}$ and by Pan and Chen for the $U_{q}(n)$
generalization of $U(n)$.$^{[31]}$
\vskip .3cm
   Based on these methods, several packages have been developed for numerically evaluating CGCs of
$U(n)$, especially of $SU(3)$. The earliest one is the well-known Akiyama-Draayer code for $SU(3)$
based on a combination of the tensor operator and infinitesimal generator methods.$^{[32-33]}$ 
Another is Chen's code for various couplings of $U(n)$ based on symmetric group techniques.$^{[30]}$
Still another is the RWC and CGC code for $SU(3)$ developed by Kaeding and Williams.$^{[34-36]}$
\vskip .3cm
   Very recently, Parkash and Sharatchandra worked out an algebraic formula for the general CGCs of
$SU(3)$.$^{[37]}$ The method used in their paper is based on a polynomial realization in Bargmann
space using generating functions, which was first studied by Shelepin and Karasev for the 
multiplicity-free case.$^{[38-39]}$ The final results are expressed in terms of a restricted sum
over 33 variables up to a normalization factor. To determine the value of a single CGC within this
formulation is not easy; neither the algebraic nor  numerical results are simple. Nevertheless, it is
the first algebraic expression for CGCs of $SU(3)$ with multiplicity. It should be noted, however,
that to extend this method to $n\geq 4$ cases will be much more complicated. Therefore, another
simpler and more direct approach to a resolution of the outer multiplicity problem for $SU(n)$ is
necessary.
\vskip .3cm
   The present paper is the first (I) in a series which has this as its objective. First of all, a
complementary group ${\cal U}(2n-2)$ realization of the Kronecker product $SU(n)\times SU(n)\downarrow
SU(n)$ is found according to the well-known Littlewood rule. The scheme gives a simple resolution of
the outer multiplicity. An analysis of the Littlewood rule is also used to derive a new multiplicity
formulae for $SU(n)$. Examples are given for the $SU(3)$ and $SU(4)$ cases which can, in principle, be
extended to $SU(n)$. A procedure for evaluating CGCs or RWCs of $SU(n)\supset SU(n-1)\times U(1)$ is
outlined which uses recoupling procedures. By using this method, $SU(n)\supset SU(n-1)\times U(1)$
RWCs or CGCs with outer multiplicity can be obtained analytically in some simple cases or numerically
in general.
Detailed results will be given for the $SU(3)$ and $SU(4)$ cases in Parts II and III of the series,
respectively. It should be noted that the RWCs or CGCs obtained in this way are orthogonal with respect to the
outer multiplicity labels and therefore the scheme that is canonical.
\newpage
\begin{centering}
{\large II. Littlewood rule and the complementary group }\\
\end{centering}
\vskip .4cm
   The Littlewood rule for determining Kronecker products of $SU(n)$ in $SU(n)\times SU(n)\downarrow
SU(n)$ is a reflection of the Shcur-Weyl duality relation between $SU(n)$ and the symmetric group
$S_{f}$. According to Schur-Weyl duality relation, an irrep $[\lambda ]$ of $SU(n)$ can also be regarded
as the same irrep of $S_{f}$ with $\sum^{n}_{i=1}\lambda_{i}=f$. Therefore, the Kronecker product of two
$SU(n)$ irreps $[\lambda ]\times[\mu ]$ in the decomposition $SU(n)\times SU(n)\downarrow SU(n)$ can be
obtained from the product of two S-functions of the corresponding symmetric groups:
\vskip .3cm
$$[\lambda ]\times [\mu ]=\sum_{\nu}\{\lambda\mu\nu\}[\nu ],\eqno(2.1)$$
\vskip .3cm
\noindent where $\{\lambda\mu\nu\}$ is the number of occurrence of $[\lambda ]$ in the product. To
determine all the irreps that appear on the rhs of (2.1), one can use the well-known Littlewood
rule:$^{[40]}$ First fill in the Young diagram $[\mu ]=[\mu_{1},\mu_{2},\cdots,\mu_{n}]$ with
$\mu_{1}$ symbols $a_{1}$ in the first row, $\mu_{2}$ symbols $a_{2}$ in the second row, $\mu_{3}$
symbols $a_{3}$ in the third row, $\cdots$, and $\mu_{n}$ symbols $a_{n}$ in the $n$th row. Then, the
final irrep denoted by Young diagram $[\nu ]$ can be obtained by augmenting the Young diagram $[\lambda
]$ with the $\mu_{1}$ $a_{1}$ symbols, $\mu_{2}$ $a_{2}$ symbols,$\cdots$, and $\mu_{n}$ $a_{n}$
symbols, respectively, in ways specified by the following three conditions:
\vskip .3cm
\noindent (a) No identical symbols should appear in the same column of the diagram.
\vskip .3cm
\noindent (b) If the $a_{1}$, $a_{2}$, $\cdots$, $a_{n}$ symbols are counted from right to left starting
at the top, then at each stage the number of $a_{1}$ symbols must not be less than the number of $a_{2}$
symbols, which must not be less than the number of $a_{3}$ symbols, and so on.
\vskip .3cm
\noindent (c) The Young diagram $[\nu ]$ obtained after the addition of each symbol must be standard,
that is, $\nu_{1}\geq \nu_{2}\geq\cdots\geq\nu_{n}$.
\vskip .3cm
   The Young diagram $[\nu ]$ filled with symbols $a_{1}$, $a_{2}$, $\cdots$, $a_{n}$ under restrictions
(a){--}(c) can be regarded as a special Weyl tableau of a unitary group. Recall some basic definitions
for Weyl tableau:  A Weyl tableau is a Young diagram with the boxes filled by a set of ordered indices
$a_{1}$, $a_{2}$, $\cdots$, $a_{n}$. The filling must be done such that:
\vskip .3cm
\noindent (i) no identical symbols should appear in the same column,
\vskip .3cm
\noindent (ii) the symbols must be in nondecreasing order from left to right in any row and in
increasing order from top to bottom in any column. 
\vskip .3cm
\noindent The one-to-one correspondence between the Gel'fand
symbol and the Weyl tableau is realized in the following way:
\vskip .3cm
$$\left(
\begin{array}{l}
 {~}[\nu ]\\
(m)
\end{array}\right)= W^{[\nu ]}=
\begin{array}{l}
{\begin{tabular}{|l|l|l|l|l|}
\hline
$f_{11}a_{1}$'s{~} &$f_{12}a_{2}$'s{~} &{$\cdots$}&$\cdots$ &$f_{1n}a_{n}$'s\\
\hline
\end{tabular}}\\
{\begin{tabular}{|l|l|l|l|}
$f_{22}a_{2}$'s &$f_{23}a_{3}$'s &{$\cdots$} &$f_{2n}a_{n}$'s\\
\hline
\end{tabular}}\\
{\begin{tabular}{|l|}
$\cdots\cdots\cdots\cdots$\\
\hline
\end{tabular}}\\
{\begin{tabular}{|l|}
$f_{nn}a_{n}$'s\\
\hline
\end{tabular}}
\end{array}\eqno(2.2)
$$
\vskip .3cm
\noindent where 
\vskip .3cm
$$f_{1k}=m_{1k}-m_{1k-1},~~~f_{2k}=m_{2k}-m_{2k-1},\cdots,$$

$$f_{k-1k}=m_{k-1k}-m_{k-1k-1},~~f_{kk}=m_{kk}.\eqno(2.3)$$
\vskip .3cm
\noindent In other words, a Weyl tableau $W^{[\nu ]}$ filled with $a_{1}$, $a_{2}$,$\cdots$, $a_{n}$,
corresponds to the $n$ partitions, $[\nu ](=[m_{in}]),~[m_{in-1}],\cdots,~[m_{i2}]$ and $[m_{11}]$ of a
Gel'fand symbol, where $[m_{ik}]$ is the Young diagram resulting from deleting all the boxes in the Weyl
tableau occupied by the symbols $a_{n},~a_{n-1},~\cdots,~a_{k+1}$.
\vskip .3cm
\noindent  It is clear that the definitions of the Weyl tableau and the rules for placing symbols in a 
Young diagram given by Littlewood are the same eccept for some of the restrictions given by (b). that
is, $a_{k}$'s can appear in the $i$th rows with $i<k$, and the number of $a_{k}$'s can be greater than
that of $a_{i}$'s with $i<k$ from right to left and from top to bottom, while these cases are forbidden
by the restriction (b) of the Littlewood rule. Therefore, it is obvious that the Littlewood rule for
placing symbols in a Young diagram can be regarded as a special Weyl tablueau for a unitary group. 
Hence, under the restrictions of Littlewood rule given by (b), one obtains a special Gel'fand basis of a
corresponding unitary group, which is called the complementary group for Kronecker products of $SU(n)$.
\vskip .3cm
   Assume the irrep $[\lambda ]$ has $p_{1}$ rows, while $[\mu ]$ has $p_{2}$ rows. Then, the final
irrep $[\nu ]$ has at most $p_{1}+p_{2}$ rows with $p_{1}+p_{2}\leq n$. Therefore, the complementary
group corresponding to the Kronecker product of $SU(n)$ is ${\cal U}(p_{1}+p_{2})$. A general $SU(n)$
irrep has at most $n-1$ rows because one can always use the equivalence condition $[m_{1n}m_{2n}\cdots
m_{nn}]=[m_{1n}-m_{nn}, m_{2n}-m_{nn},\cdots, m_{1n-1}-m_{nn}]$ to remove the $n$th row if it exists.
From this it follows that the minimum complementary group is ${\cal U}(2n-2)$ for general Kronecker
products of $SU(n)$.
\vskip .3cm
   Using the correspondence between Weyl tableau and a Gel'fand symbol, one can easily find the
following relations among coupled and uncoupled state labels of ${\cal U}(2n-2)$.
\vskip .3cm
$$
\begin{array}{l}
{\begin{tabular}{|l|}
\hline
$~~~~~~~\lambda_{1}~~~~~~~~$\\
\hline
\end{tabular}}\\
{\begin{tabular}{|l|}
$~~~~~\lambda_{2}~~~~~~$\\
\hline
\end{tabular}}\\
{\begin{tabular}{|l|}
$\cdots\cdots~~~$\\
\hline
\end{tabular}}\\
{\begin{tabular}{|l|}
$\lambda_{n-1}$'s\\
\hline
\end{tabular}}
\end{array}\Rightarrow
\left(
\begin{array}{l}
[\lambda_{1}\lambda_{2}\cdots\lambda_{n-1}\dot{0}]~~{\cal U}(2n-2)\\
~~~~~~~~~~\cdots\cdots\\
{[\lambda_{1}\lambda_{2}\cdots\lambda_{n-1}\dot{0}]~~{\cal U}(n-1)}\\
~~~~~~~~~~\rho\\
\end{array}\right)
\eqno(2.4a)
$$   
\vskip .3cm
$$
\begin{array}{l}
{\begin{tabular}{|l|}
\hline
$~~~~~~~\mu_{1}~ a_{1}$'s~~~~~~~~~~\\
\hline
\end{tabular}}\\
{\begin{tabular}{|l|}
$~~~~~\mu_{2}~ a_{2}$'s~~~~~\\
\hline
\end{tabular}}\\
{\begin{tabular}{|l|}
$~~~\cdots\cdots~~~~$\\
\hline
\end{tabular}}\\
{\begin{tabular}{|l|}
$\mu_{n-1}~a_{n}$'s\\
\hline
\end{tabular}}
\end{array}\Rightarrow
\left(
\begin{array}{l}
[\mu_{1}\mu_{2}\cdots\mu_{n-1}\dot{0}]~~{\cal U}(2n-2)\\
{[\mu_{1}\mu_{2}\cdots\mu_{n-2}\dot{0}]~~{\cal U}(2n-3)}\\
~~~~~~~~~~\cdots\cdots\\
{~~~~~~[\mu_{1}\mu_{2}\dot{0}]~~~~~~~~{\cal U}(n+1)}\\
{~~~~~~~~[\mu_{1}\dot{0}]~~~~~~~~~~~~{\cal U}(n)}\\
{~~~~~~~~~~[\dot{0}]~~~~~~~~~~~{\cal U}(n-1)}
\end{array}\right)
\eqno(2.4b)
$$
\noindent while the final coupled ${\cal U}(2n-2)$ basis is
\vskip .3cm
$$\left(
\begin{array}{l}
{[\nu_{1}\nu_{2}\cdots\nu_{n}\dot{0}]~(\tau)~~{\cal U}(2n-2)}\\
\\
~~~~~~~~~~~~~~(\tau)\\
\\
{[\lambda_{1}\lambda_{2}\cdots\lambda_{n-1}\dot{0}]~~~{\cal U}(n-1)}\\
~~~~~~~~~\rho
\end{array}
\right),\eqno(2.4c)$$
\vskip .3cm
\noindent where $(\tau)$ stands for intermediate sublabels between ${\cal U}(2n-2)$ and ${\cal
U}(n-1)$ given by the Littlewood rule, which is simultaneously the outer multiplicity label of both
${\cal U}(2n-2)$ and $SU(n)$, and $\rho$ represents sublabels of ${\cal U}(n-1)$.
\vskip .3cm
   Therefore, $(\tau)$ can be regarded as  multiplicity labels of $SU(n)$. For example,
the final coupled state can be written as
\vskip .3cm
$$\left( 
\begin{array}{l}
{[\nu_{1}\nu_{2}\cdots\nu_{n}]~(\tau)}\\
~~~~~~~(\nu )
\end{array}\right),\eqno(2.5)$$
\vskip .3cm
\noindent where $(\nu )$ stands for sublabels of $SU(n)$. Expression (2.5) is similar to the upper Gel'fand
pattern introduced by Biedenharn et al.$^{[1-8]}$ The final coupled ${\cal U}(2n-2)$ labels $(\tau )$ in
(2.4c) provide the outer multiplicity labels needed in the decomposition $[\lambda ]\times [\mu ]\downarrow
[\nu ]$. This will be discussed further in the next section. 
\vskip .6cm
\begin{centering}
{\large III. Outer multiplicity problem of SU(3) and SU(4) }\\
\end{centering}
\vskip .4cm
   As noted above, the outer multiplicity in the decomposition of the Kronecker products of $SU(n)\times
SU(n)\downarrow SU(n)$ is the main obstacle in applications of algebraic methods to physical problems. 
There are a lot of articles devoted to this subject. In order to resolve the problem for the $SU(3)$,
Hecht$^{[22]}$ proposed an external labeling operator of third order, an operator that may be related to
the one proposed by Moshinsky$^{[16]}$ in terms of the complementary $U(4)\supset U(2)\times U(2)$
chain. Alisauskas and Kulish$^{[41]}$ have also proposed an external labeling operator, a fourth order
form suggested by Sharp$^{[42]}$ in a study of Yang-Baxter equations. There are also other articles on
this subject. For example, new Casimir operators, the so called chiral Casimirs, were introduced in
[16, 43-44]. Also, various formulae$^{[19, 45-48]}$ for the multiplicity of $SU(3)$ exist in the
literature, however, such expressions are normally not linked to the $SU(3)$ coupling and recoupling
coefficients problem. There is still no general formula for the outer multiplicity of $SU(n)$ with
$n\geq 4$. In this article and forthcoming papers, the complementary group ${\cal U}(2n-2)$ to the
$SU(n)\times SU(n)\downarrow SU(n)$ will be shown to be a powerful tool for deriving both multiplicity
formulae and coupling and recoupling coefficients of $SU(n)$. Multiplicity formulae for $SU(3)$ and
$SU(4)$ are considered below.
\vskip .4cm
\noindent {\bf (1) SU(3) case.} Consider the general Keronecker product
$(\lambda_{1}\mu_{1})\times(\lambda_{2}\mu_{2})$, where the well-known notation for $SU(3)$ in physics
is adopted. The irrep $(\lambda\mu )$ can be expressed in terms of a two-rowed Young diagram
$[\nu_{1}\nu_{2}]$ with $\nu_{1}=\lambda +\mu$, and $\nu_{2}=\mu$. Using the Littlewood rule, the
decomposition of $(\lambda_{1}\mu_{1})\times(\lambda_{2}\mu_{2})$ can be expressed in terms of a
quintuple sum.
\vskip .3cm
$$(\lambda_{1}\mu_{1})\times(\lambda_{2}\mu_{2})=\sum^{\lambda_{2}+\mu_{2}}_{k_{1}=0}
\sum^{\min(\lambda_{1},~\lambda_{2}+\mu_{2}-k_{1})}_{k_{2}=0}
\sum^{\min(\mu_{1},~\lambda_{2}+\mu_{2}-k_{1}-k_{2})}_{k_{3}=0}
\sum^{\min(\lambda_{1}+k_{1}-k_{2},~\mu_{2},~k_{1})}_{n_{1}=0}\times$$

$$\sum^{\min(\mu_{2}-n_{1},~\mu_{1}+k_{2}-k_{3},~k_{1}+k_{2}-n_{1})}_{n_{2}=0}
[\lambda_{1}+\mu_{1}+k_{1},~\mu_{1}+k_{2}+n_{1},~k_{3}+n_{2}],\eqno(3.1)$$
\vskip .3cm
\noindent where the constraints 
\vskip .3cm
$$\sum^{3}_{i=1}k_{i}=\lambda_{2}+\mu_{2},~~\sum^{2}_{i=1}n_{i}=\mu_{2}\eqno(3.2)$$
\vskip .3cm
\noindent apply in the summation. Expression (3.1) can be further simplified, for example,
to O'Reilly's formula$^{[47]}$ in which only a triple sum appears. However, (3.1) can be used to help
determine a multiplicity formula and determine the multiplicity labels of the complementary group.
\vskip .3cm
   Consider a Young diagram of the resultant irrep $[m_{1}m_{2}m_{3}]$ according to (3.1) with
conditions given by (3.2):
\vskip .3cm
$$\begin{array}{l}
{\begin{tabular}{|l|l|}
\hline
$~~~~~~~~~~~~~~~~~~~~~\lambda_{1}+\mu_{1}~~~~~~~~~~~~~~~~$  &~~~~~$k_{1}~~~~~^{\alpha}$\\
\hline
\end{tabular}}\\
{\begin{tabular}{|l|l|l|l|}
$~~~~~~~~~\mu_{1}~~~~~~~~~~$ &$m_{2}-\mu_{1}-\eta~~^{\alpha}$  &$~\eta~~~^{\beta}$\\
\hline
\end{tabular}}\\
{\begin{tabular}{|l|l|}
$m_{3}-\mu_{2}+\eta~~^{\alpha}$ &$\mu_{2}-\eta~~^{\beta}$\\
\hline
\end{tabular}}
\end{array}\eqno(3.3)
$$
\vskip .3cm
\noindent where $\alpha$ and $\beta$ are the $a_{i}$ symbols of the Littlewood rule for $SU(3)$. The
labels in (3.3) have been arranged to acommodate the constraints of (3.2) and to yield a multiplicity
formula very easily. In this forms it is obvious that a diagram with the same number of boxes in each
row can only appear repeatedly when $\eta$ is not a fixed integer. Therefore, $\eta$ can be regarded as
the multiplicity label of SU(3).
\vskip .3cm
   According to Littlewood rule (a){--}(c), it is easy to derive the following limits on $\eta$:
\vskip .3cm
$$\eta_{\min}\leq\eta\leq\eta_{\max},\eqno(3.4)$$
\vskip .3cm
\noindent where
\vskip .3cm
$$\eta_{\max}=\min(m_{1}-\lambda_{1}-\mu_{1},~\mu_{2},~m_{2}-\mu_{1},~\lambda_{2}+\mu_{2}-m_{3},~
\mu_{1}+\mu_{2}-m_{3},~m_{2}-m_{3}),$$
$$\eta_{\min}=\max( 0,~\mu_{2}-m_{3},~m_{2}-\lambda_{1}-\mu_{1}).\eqno(3.5)$$
\vskip .3cm
\noindent Hence, the multiplicity of $[m_{1}m_{2}m_{3}]\equiv (m_{1}-m_{2},~m_{2}-m_{3})$ occurring
in the Kronecker product $(\lambda_{1}\mu_{1})\times (\lambda_{2}\mu_{2})$ is given by
\vskip .3cm
$${\rm Multi}(SU_{3})=\eta_{\max}-\eta_{\min}+1.\eqno(3.6)$$
\vskip .3cm
\noindent This expression is very simple and more transparent than others found in the literature.
\vskip .3cm
   In this case, the complementary group is ${\cal U}(4)$. The Gel'fand symbol of ${\cal U}(4)$
corresponding to the resultant irrep of $SU(3)$ given in (3.3) is
\vskip .3cm
$$\left(
\begin{array}{l}
{~~~~~~~~~~~[m_{1}m_{2}m_{3}0]~\eta}\\
{[m_{1}~m_{2}-\eta~~m_{3}-\mu_{2}+\eta]}\\
{~~~~~~~~~~[\lambda_{1}+\mu_{1}~\mu_{1}]}\\
~~~~~~~~~~~~~~~~~\rho
\end{array}
\right),~\eta =\eta_{\min},~\eta_{\min}+1,\cdots,~\eta_{\max},\eqno(3.7)$$
\vskip .3cm
\noindent where $\rho$ is the intrinsic label for ${\cal U}(1)$, which is not important for our purpose.
Some conditions in (3.5) can also be easily obtained from the betweeness conditions of the entries in
the Gel'fand symbol (3.7). However, the remaining conditions in (3.5) can only be deduced from the
Littlewood rule (b), and can not be obtained from the betweeness conditions. Hence, only one outer
multiplicity label is needed in the decomposition of $SU(3)\times SU(3)\downarrow SU(3)$. This is why
the CGCs of $SU(3)$ can be determined numerically by using only one type of tensor operator.$^{[32-36]}$
\vskip .3cm
   In contrast with the so-called canonical labeling scheme proposed by Biedenharn et al., in which
three independent shifts determined by an upper pattern are introduced, the complementary ${\cal U}(4)$
group provides only one outer multiplicity label in the $SU(3)$ case. The complementary group labeling
scheme is therefore a very economical way to label the outer multiplicity of $SU(3)$, and by extension,
of $SU(n)$.
\vskip .3cm
   Furthermore, the upper pattern labeling scheme given by Biedenharn et al is also equivalent to our
labeling scheme, which will be proved in our next paper. However, similar to the complementary group
labeling scheme, some restrictions on the ranges of $\Gamma_{22}$ in the upper pattern 
$[\Gamma_{12},~\Gamma_{22}]=[m_{1}+m_{2}-\lambda_{1}-2\mu_{1}-\Gamma_{22},~\Gamma_{22}]$ for the coupling
$(\lambda_{1}\mu_{1})\times (\lambda_{2}\mu_{2})\downarrow [m_{1}m_{2}m_{3}]$ should be  
obtained from the Littlewood rule of $SU(3)$. Actually, the ranges of $\Gamma_{22}$ should be
the same as those  of $\eta$ given by (3.5), which, however, can not be derived directly from restrictions
on upper pattern labels. For example, $[422]$ occurs only once in the decomposition $[310]\times [310]$.
However, there are two sets of upper pattern labels
\vskip .3cm
$$\left(
\begin{array}{l}
~~~\Gamma_{11}\\
\Gamma_{12}~~\Gamma_{22}\\
\end{array}\right)=\left(
\begin{array}{l}
~~1\\
2~~0\\
\end{array}\right)~;~\left(
\begin{array}{l}
~~1\\
1~~1\\
\end{array}\right)$$
\vskip .3cm
\noindent are allowed according to the upper pattern labeling scheme. Actually, the state labelled
by $\tiny\left(
\begin{array}{l}
~~1\\
1~~1\\
\end{array}\right)$
 should be eleminated according to the Littlewood rule. Therefore, restrictions from Littlewood rule must
apply to the upper pattern labeling scheme, which was not mentioned in their papers [1-8], and indeed difficult
to be obtained directly from their methods.

\vskip .3cm
\noindent {\bf (2) SU(4) case.} A general $SU(4)$ irrep has three rows. Using the Littlewood rule, the
following formula for the decomposition of $SU(4)\times SU(4)\downarrow SU(4)$ can be determined:
\vskip .3cm
$$[\lambda_{1}\lambda_{2}\lambda_{3}]\times [\mu_{1}\mu_{2}\mu_{3}]=\sum^{\mu _{1}}_{k_{1}=0}
\sum^{\min(\mu_{1}-k_{1},\lambda_{1}-\lambda_{2})}_{k_{2}=0}~
\sum^{\min(\lambda_{2}-\lambda_{3},\mu_{1}-k_{1}-k_{2})}_{k_{3}=0}
~\sum^{\min(\lambda_{3},\mu_{1}-k_{1}-k_{2}-k_{3})}_{k_{4}=0}\times$$

$$\sum^{\min(\lambda_{1}+k_{1}-\lambda_{2}-k_{2},\mu_{2},k_{1})}_{l_{1}=0}~
\sum^{\min(\lambda_{2}+k_{2}-\lambda_{3}-k_{3},\mu_{2}-l_{1},k_{1}+k_{2}-l_{1})}_{l_{2}=0}\times$$

$$\sum^{\min(\lambda_{3}+k_{3}-k_{4},\mu_{2}-l_{1}-l_{2},k_{1}+k_{2}+k_{3}-l_{1}-l_{2})}_{l_{3}=0}~
\sum^{\min(\lambda_{2}+k_{2}+l_{1}-\lambda_{3}-k_{3}-l_{2},\mu_{3},l_{1})}_{n_{1}=0}\times$$

$$\sum^{\min(\lambda_{3}+k_{3}+l_{2}-k_{4}-l_{3},\mu_{3}-n_{1},l_{1}+l_{2}-n_{1})}_{n_{2}=0}
[\lambda_{1}+k_{1},~\lambda_{2}+k_{2}+l_{1},~\lambda_{3}+k_{3}+l_{2}+n_{1},~k_{4}+l_{3}+n_{2}],
\eqno(3.8)$$
\vskip .3cm
\noindent where the following constraints 
\vskip .3cm
$$\sum^{4}_{i=1}k_{i}=\mu_{1},~~\sum^{3}_{i=1}l_{1}=\mu_{2},~~\sum^{2}_{i=1}n_{i}=\mu_{3}\eqno(3.9)$$
\vskip .3cm
\noindent apply in the summation. In the resultant irrep $[\nu_{1}\nu_{2}\nu_{3}\nu_{4}]\equiv
[\lambda_{1}+k_{1},~\lambda_{2}+k_{2}+l_{1},~\lambda_{3}+k_{3}+l_{2}+n_{1},~k_{4}+l_{3}+n_{2}]$ with
the restrictions given by (3.9) there may be six ways to relabel the configuration which leave the
irrep unchanged: 
\vskip .3cm
$$
\begin{array}{l}
{\begin{tabular}{|l|l|}
\hline
$~~~~~~~~~~~~~~~~\lambda_{1}~~~~~~~~~~~~$ &~~~~$k_{1}~~~^{\alpha}$\\
\hline
\end{tabular}}\\
{\begin{tabular}{|l|l|l|l|}
$~~~~~~~~\lambda_{2}~~~~~~~~~$ &$k_{2}~~^{\alpha}$  &$l_{1}~~^{\beta}$\\
\hline
\end{tabular}}\\
{\begin{tabular}{|l|l|l|l|}
$~~~\lambda_{3}~~~$ &$k_{3}~~^{\alpha}$ &$l_{2}~~^{\beta}$ &$n_{1}~~^{\gamma}$\\
\hline
\end{tabular}}\\
{\begin{tabular}{|l|l|l|}
$k_{4}~~^{\alpha}$ &$l_{3}~~^{\beta}$ &$n_{2}~~^{\gamma}$\\
\hline
\end{tabular}}
\end{array}\eqno(3.10a)
$$
\vskip .3cm
\noindent where 
\vskip .3cm
$$k_{1}=\nu_{1}-\lambda_{1},$$

$$k_{2}=\nu_{2}-\lambda_{2}-\xi_{1}-\xi_{2},$$

$$k_{3}=\nu_{3}-\lambda_{3}-\mu_{2}+\xi_{1}-\xi_{3}-\xi_{5}-\xi_{6},$$

$$k_{4}=\nu_{4}-\mu_{3}+\xi_{2}+\xi_{3}+\xi_{5}+\xi_{6},$$

$$l_{1}=\xi_{1}+\xi_{2},$$

$$l_{2}=\mu_{2}-\xi_{1}-\xi_{4}+\xi_{5},$$

$$l_{3}=\xi_{4}-\xi_{2}-\xi_{5},$$

$$n_{1}=\xi_{3}+\xi_{4}+\xi_{6},$$

$$n_{2}=\mu_{3}-\xi_{3}-\xi_{4}-\xi_{6},\eqno(3.10b)$$
\vskip .3cm
\noindent and $\alpha$, $\beta$, and $\gamma$ are the symbols filling in each box according to the
Littlewood rule. However, by using the following transformation
\vskip .3cm
$$\eta_{1}=\xi_{1}+\xi_{2},~~\eta_{2}=\xi_{3}+\xi_{4}+\xi_{6},~~\eta_{3}=\xi_{4}-\xi_{2}-\xi_{5},\eqno(3.11)$$
\vskip .3cm
\noindent it can be shown that only three variables $\eta_{i}$ with $i=1,~2,$ and $3$ are independent.
Therefore, (3.10) can be relabelled in terms of these 3 variables,
\vskip .3cm
$$k_{1}=\nu_{1}-\lambda_{1},$$

$$k_{2}=\nu_{2}-\lambda_{2}-\eta_{1},$$

$$k_{3}=\nu_{3}-\lambda_{3}-\mu_{2}+\eta_{1}-\eta_{2}+\eta_{3},$$

$$k_{4}=\nu_{4}-\mu_{3}+\eta_{2}-\eta_{3},$$

$$l_{1}=\eta_{1},$$

$$l_{2}=\mu_{2}-\eta_{1}-\eta_{3},$$

$$l_{3}=\eta_{3},$$

$$n_{1}=\eta_{2},$$

$$n_{2}=\mu_{3}-\eta_{2}.\eqno(3.12)$$

\vskip .3cm
  Applying the Littlewood rule to this result yields the following boundary conditions for the outer
multiplicity labels $\eta_{1}$, $\eta_{2}$, and $\eta_{3}$.
\vskip .3cm
 $$\eta_{1\min}\leq\eta_{1}\leq\eta_{1\max},~~\eta_{2\min}\leq\eta_{2}\leq\eta_{2\max},~~
\eta_{3\min}\leq\eta_{3}\leq\eta_{3\max},\eqno(3.13)$$
\vskip .3cm
\noindent where
\vskip .3cm
$$\eta_{1\min}=\max(0,~\nu_{2}-\lambda_{1}),~~\eta_{1\max}=\min(\nu_{2}-\lambda_{2},~\nu_{1}-\lambda_{1}),$$

$$\eta_{2\min}=\max(\nu_{3}-\nu_{2}+\eta_{1},~0),~~\eta_{2\max}=\min(\eta_{1},~\mu_{3},~\nu_{3}-\nu_{4}),$$

$$\eta_{3\min}=\max(2\eta_{2}-\eta_{1}+\mu_{2}+\nu_{4}-\nu_{3}-\mu_{3},~\eta_{2}-\eta_{1}+\lambda_{3}+\mu_{2}
-\nu_{3},\eta_{2}+\nu_{4}-\lambda_{3}-\mu_{3},$$

$$~\eta_{2}+\lambda_{1}+\lambda_{2}+\lambda_{3}+2\mu_{2}-\nu_{1}-\nu_{2}-\nu_{3},
~\eta_{1}+\lambda_{1}+\lambda_{2}+\mu_{2}-\nu_{1}-\nu_{2},$$

$$0,~{\rm Int}[(\eta_{2}+\lambda_{1}+\lambda_{2}+\lambda_{3}+2\mu_{2}-\nu_{1}-\nu_{2}-\nu_{3})/2]),$$

$$\eta_{3\max}=\min(\mu_{2}-\eta_{1},~\nu_{4}-\mu_{3}+\eta_{2},~\mu_{2}-\mu_{3},
~\lambda_{2}-\nu_{3}+\mu_{2}-\eta_{1}+\eta_{2},~\mu_{2}-\eta_{2}),\eqno(3.14)$$
\vskip .3cm
\noindent where ${\rm Int}[x]$ is the integer part of $x$.
Thus, the multiplicity of $[\nu_{1},~\nu_{2},~\nu_{3},~\nu_{4}]\equiv
[\nu_{1}-\nu_{4},~\nu_{2}-\nu_{4},~\nu_{3}-\nu_{4}]$ appearing in the Kronecker product 
$[\lambda_{1}\lambda_{2}\lambda_{3}]\times [\mu_{1}\mu_{2}\mu_{3}]$ can be calculated by
\vskip .3cm
$${\rm Multi}(SU_{4})=
\sum^{\eta_{1\max}}_{\eta_{1}=\eta_{1\min}}~\sum^{\eta_{2\max}(\eta_{1})}_{\eta_{2}=
\eta_{2\min}(\eta_{1})}~\sum^{\eta_{3\max}(\eta_{1},~\eta_{2})}_{\eta_{3}=
\eta_{3\min}(\eta_{1},~\eta_{2})}.
\eqno(3.15)$$
\vskip .3cm
   The complementary group of the Kronecker product $[\lambda_{1}\lambda_{2}\lambda_{3}]\times
[\mu_{1}\mu_{2}\mu_{3}]$ of $SU(4)$ is
${\cal U}(6)$ with the following special Gel'fand labels
\vskip .3cm
$$\left(
\begin{array}{l}
{~~~~~~~~~~~~~~~~~~~~~~~~~[\nu_{1}~\nu_{2}~\nu_{3}~\nu_{4}]~(\eta_{1}\eta_{2}\eta_{3})~~~~~~~~~~~~~~~~~~~~~~{\cal
U}(6)}\\
{~~~~~~~~~~~~~~~~[\nu_{1},~\nu_{2},~\nu_{3}-\eta_{2},~\nu_{4}-\mu_{3}+
\eta_{2},~0]~~~~~~~~~~~~~~~{\cal U}(5)}\\
{[\nu_{1},~\nu_{2}-\eta_{1},~\nu_{3}-\mu_{2}+\eta_{1}+\eta_{3}-\eta_{2},
~\nu_{4}-\mu_{3}+\eta_{2}-\eta_{3}]~~ {\cal U}(4)}\\ 
{~~~~~~~~~~~~~~~~~~~~~~~~~~~[\lambda_{1}~\lambda_{2}~\lambda_{3}]~~~~~~~~~~~~~~~~~~~~~~~~~~~~~~~~~~{\cal
U}(3)}\\
 ~~~~~~~~~~~~~~~~~~~~~~~~~~~~~~~~~\rho
\end{array}\right).
\eqno(3.16)$$
\vskip .3cm
\noindent The Gel'fand labels of
$[\lambda_{1}\lambda_{2}\lambda_{3}\dot{0}]$ and
$[\mu_{1}\mu_{2}\mu_{3}\dot{0}]$ for ${\cal U}(6)$ are
\vskip .3cm
$$\left(
\begin{array}{l}
{~[\lambda_{1}\lambda_{2}\lambda_{3}\dot{0}]~~~~~~{\cal U}(6)}\\
{~[\lambda_{1}\lambda_{2}\lambda_{3}\dot{0}]~~~~~~{\cal
U}(5)}\\
{~[\lambda_{1}\lambda_{2}\lambda_{3}0]~~~~~~{\cal U}(4)}\\
{~~[\lambda_{1}\lambda_{2}\lambda_{3}]~~~~~~{\cal U}(3)}\\
~~~~~~\rho
\end{array}\right)~~~{\rm and~}~~
\left(
\begin{array}{l}
{~[\mu_{1}\mu_{2}\mu_{3}\dot{0}]~~~~~~{\cal U}(6)}\\
{~~~[\mu_{1}\mu_{2}\dot{0}]~~~~~~~{\cal
U}(5)}\\
{~~~~~[\mu_{1}0]~~~~~~~~{\cal U}(4)}\\
{~~~~~~~[\dot{0}]~~~~~~~~~{\cal U}(3)}\\
\end{array}\right).
\eqno(3.17)$$
\vskip .3cm
\noindent Again, most of the boundary conditions for the multiplicity labels $\eta_{1},~\eta_{2}$, and
$\eta_{3}$ can be obtained from the betweeness conditions for the Gel'fand symbol shown in (3.16).
However, the remaining conditions can only be deduced from the Littlewood rule because (3.16) is a
special Gel'fand basis for the canonical chain ${\cal U}(6)\supset{\cal U}(5)\supset\cdots\supset{\cal
U}(2)\supset {\cal U}(1)$. From this development it is clear that there are at most 3 quantum numbers
needed to label the outer multiplicity for the decomposition $SU(4)\times SU(4)\downarrow SU(4)$. In the
canonical unit tensor approach proposed by Biedenharn et al. for the $SU(4)$ case, there are 4 shifts
out of 6 upper labels, of which only 3 labels are independent.$^{[1-8]}$ Similar to $SU(3)$ case, restrictions
from Littlewood rule  must apply to eliminate superfluous multiplicity states in the upper pattern
labeling scheme.
\vskip .3cm
   It should be noted that any $SU(n)$ function, for example, CGCs, RWCs, or Racah coefficients, etc.,
is rank $n$ independent, and only depends on boxes contained in the Young diagrams of the corresponding
irreps because of the Schur-Weyl duality relation between $SU(n)$ and $S_{f}$. For example, the
multiplicity of Kronecker product for two two-rowed irreps of $SU(n)$ is the same as that of $SU(3)$,
and that for two three-rowed irreps is the same as that of $SU(4)$, and so on. Hence, the results for
$SU(3)$ and $SU(4)$ apply for the general $SU(n)$ case as well.  As a trivial example, note that the
$SU(3)$ multiplicity expression follows from the one for $SU(4)$ in the two-rowed limit ($\eta_{1}
\rightarrow \eta$, $\eta_{2}=0$ and $\eta_{3}=0$) of the theory though this fact can not be clearly
seen from (3.14). 
\vskip .6cm
\begin{centering}
{\large IV. Conclusions }\\
\end{centering}
\vskip .4cm
 In this paper, a complementary group $U(2n-2)$ to $SU(n)$ is found that gives a complete realization of all
the features of the Littlewood rule in the Kronecker product decomposition of $SU(n)\times SU(n)\downarrow
SU(n)$. By using this scheme, the outer multiplicity labels for $SU(n)$ can be easily assigned, being
nothing other than a set of sublabels of the special Gel'fand basis of the complementary ${\cal U}(2n-2)$
group. Furthermore, within this framework, most of the boundary conditions on the multiplicity labels can
be easily obtained from the betweenness conditions of the Gel'fand symbols of ${\cal U}(2n-2)$, while the
remaining conditions must be deduced from the Littlewood rule. The method was used to obtained simple
multiplicity formulae for $SU(3)$ and $SU(4)$. In addition, in the coupling of two $SU(n)$ irreps, the basis
for $SU(n)$ can further be labeled by the final ${\cal U}(2n-2)$ sublabels $\eta_{i}$ obtained from the
coupling of two uncoupled basis vectors of the corresponding specail Gel'fand basis of 
${\cal U}(2n-2)$, which are missing within the $SU(n)$ group. This situation is very similar to that of  the
canonical unit tensor approach proposed by Biedenharn et al. However, in the canonical unit tensor
approach, there are $n$ independent shifts indicated by the  upper pattern of $U(n)$ from the $n(n-1)/2$
upper labels. While these upper indices can be used to label the outer multiplicity of $U(n)$, there
may very well be superfluous degree-of-freedom among the labels and these may be eliminated, especially,
restrictions from Littlewood rule of $SU(n)$ must apply to eliminate superfulous multiplicity states which
are not allowed in the decomposition.
\vskip .3cm
   It should be stated that the same complementary group to the resolution of $SU(n)$ was also considered
in [15]. However, the method used and the final outcome are all different. Firstly, In [15], this
complementary group was derived by using boson realizations given by Moshinsky.$^{[16]}$ While it now comes
naturally from the Littlewood rule. Secondly, according to [15], the complementary group should be labeled
in terms of a noncanonical chain ${\cal U}(2n-2)\supset{\cal U}(n-1)\times{\cal U}(n-1)$. In this way, the
RWCs of $SU(n)$ still can not easily be derived because  new  inner multiplicity occurs in the decomposition
${\cal U}(2n-2)\downarrow{\cal U}(n-1)\times{\cal U}(n-1)$. In order to overcome this difficulty, another
kind of Wigner coefficients, the so called auxiliary Wigner coefficients was defined in [15], which is
different from the standard definition of WCs, and satisfy  another type of orthogonality conditions.
We shall discuss these special WCs in the next paper. It shall be show in the next paper that one can derive
analytical expressions in some simple cases and the corresponding algorithms for
$SU(n)$ RWCs or CGCs with multiplicity in general in both the canonical and noncanonical bases within
this labeling scheme if the multiplicity-free coefficients in these bases are known. 
\vskip .3cm
   To reiterate an important
point, the complementary group
${\cal U}(2n-2)$ scheme for labeling outer multiplicities in Kronecker products of
$SU(n)$ is itself a canonical scheme because the basis of $SU(n)$ labeled in this way is orthogonal with
respect to the outer multiplicity labels. A general procedure for evaluating RWCs or CGCs for $SU(3)$
and $SU(4)$ will be given in the forthcoming papers.
\vskip .8cm
\noindent {\bf Acknowledgment}
\vskip .5cm
   The project was supported in part by the US National Science Foundation and the State Education
Commission of China.
   
\vskip .5cm
\begin{tabbing}
\=1111\=22222222222222222222222222222222222222222222222222222222222222222222222222222222\=\kill\\
\>{[1]}\>{G. E. Baird and L. C. Biedenharn, J. Math. Phys., {\bf 4} (1963) 1449; {\bf 5} (1964) 1723;}\\
\>{}\>{{\bf 5} (1964) 1730; {\bf 6} (1965) 1847}\\
\>{[2]}\>{L. C. Biedenharn, A. Giovannini, and J. D. Louck, J. Math. Phys., {\bf 8} (1967) 691}\\
\>{[3]}\>{L. C. Biedenharn and J. D. Louck, Commun. Math. Phys., {\bf 8} (1968) 89}\\
\>{[4]}\>{L. C. Biedenharn, J. D. Louck, E. Chacon, and M. Ciftan, J. Math. Phys., {\bf 13} (1972)
1957}\\
\>{[5]}\>{E. Chacon, M. Ciftan, and L. C. Biedenharn, J. Math. Phys., {\bf 13} (1972) 577}\\
\>{[6]}\>{L. C. Biedenharn and J. D. Louck, Commun. Math. Phys., {\bf 93} (1984) 143}\\
\>{[7]}\>{L. C. Biedenharn, M. A. Lohe, and J. D Louck, J. Math. Phys., {\bf 26} (1985) 1458}\\
\>{[8]}\>{L. C. Bidenharn, M. A. Lohe, and H. T. Williams, J. Math. Phys., {\bf 35} (1994) 6072}\\
\>{[9]}\>{R. Le Blanc and K. T. Hecht, J. Phys. A, {\bf 20} (1987) 4613}\\
\>{[10]}\>{R. Le Blanc and L. C. Biedenharn, J. Phys. A, {\bf 22} (1989) 4613}\\
\>{[11]}\>{K. T. Hecht and L. C. Biedenharn, J. Math. Phys., {\bf 31} (1990) 2781}\\
\>{[12]}\>{S. J. Ali\v{s}auskas, A. A. Jucy, and A. P. Jucy, J. Math. Phys., {\bf 13} (1972) 1329}\\
\>{[13]}\>{S. J. Ali\v{s}uaskas, V. V. Vanagas, and J. P. Jucy, Dokl. Akad. Nauk. SSSR, {\bf 197} (1971)
804}\\
\>{[14]}\>{S. J. Ali\v{s}uaskas, Sov. J. Part. Nucl., {\bf 14} (1983) 563}\\
\>{[15]}\>{T. A. Brody, M. Moshinsky, and I. Renero, J. Math. Phys., {\bf 6} (1965) 1540}\\
\>{[16]}\>{M. Moshinsky, J. Math. Phys., {\bf 4} (1963) 1128; Rev. Mod. Phys., {\bf 34} (1962) 813}\\
\>{[17]}\>{J. R. Derome and W. T. Sharp, J. Math. Phys., {\bf 7} (1966) 612}\\
\>{[18]}\>{J. R. Derome, J. Math. Phys., {\bf 8} (1967) 714}\\
\>{[19]}\>{M. Resnikoff, J. Math. Phys., {\bf 8} (1967) 63}\\
\>{[20]}\>{Z. Pluha\v{r}, Yu F. Smirnov, and V. N. Tolstoy, J. Phys. A,{\bf 19} (1986) 21}\\
\>{[21]}\>{Z. Pluha\v{r}, L. J. Weigert, and P. Holan, J. Phys. A, {\bf 19} (1986) 29}\\
\>{[22]}\>{K. T. Hecht, Nucl Phys, {\bf 62} (1965) 1}\\
\>{[23]}\>{A. U. Klimyk and A. M. Gavrilik, J. Math. Phys., {\bf 20} (1979) 1624}\\
\>{[24]}\>{R. Le Blanc and D. J. Rowe, J. Phys. A, {\bf 19} (1986) 2913}\\
\>{[25]}\>{S. J. Ali\v{s}auskas, J. Math. Phys., {\bf 29} (1988) 2351}\\
\>{[26]}\>{S. J. Ali\v{s}auskas, J. Math. Phys., {\bf 31} (1990) 1325}\\
\>{[27]}\>{S. J. Ali\v{s}auskas, J. Math. Phys., {\bf 33} (1992) 1983}\\
\>{[28]}\>{S. J. Ali\v{s}auskas, J. Phys. A, {\bf 29} (1996) 2687}\\
\>{[29]}\>{P. Kramer, Z. Phys., {\bf 216} (1968) 68; {\bf 205} (1967) 181}\\
\>{[30]}\>{J. Q. Chen, P. N. Wang, Z. M. L\"{u}, and X. B. Wu, Tables of the CG, Racah, and}\\
\>{}\>{subduction coefficients of $SU(n)$ groups (Singapore, World Scientific, 1987)}\\
\>{[31]}\>{Feng Pan and J. Q. Chen, J. Phys. A, {\bf 26} (1993) 4299; J. Math. Phys. {\bf 34}}\\
\>{}\>{(1993) 4305; 4316}\\
\>{[32]}\>{J. P. Draayer and Y. Akiyama, J. Math. Phys. {\bf 14} (1973) 1904}\\
\>{[33]}\>{Y. Akiyama and J. P. Draayer, Comp. Phys. Commun., {\bf 5} (1973) 405}\\
\>{[34]}\>{T. A. Kaeding, Comp. Phys. Commun., {\bf 85} (1995) 82}\\
\>{[35]}\>{T. A. Kaeding and H. T. Williams, Commp. Phys. Commun., {\bf 98} (1996) 398}\\
\>{[36]}\>{H. T. Williams, J. Math. Phys., {\bf 37} (1996) 4187}\\
\>{[37]}\>{J. S. Prakash and H. S. Sharatchandra, J. Math. Phys., {\bf 37} (1996) 6530}\\
\>{[38]}\>{L. A. Shelepin and V. P. Karasev, Sov. J. Nucl. Phys., {\bf 5} (1967) 156}\\
\>{[39]}\>{V. P. Karasev and L. A. Shelepin, Sov. J. Nucl. Phys., {\bf 7} (1968) 678}\\
\>{[40]}\>{D. E. Littlewood, The Theory of Gourp Characters, 2nd edn. (Oxford, Claredom, 1950)}\\
\>{[41]}\>{S. J. Ali\v{s}auskas and P. P. Kulish, J. Sov. Math., {\bf 35} (1986) 2653}\\
\>{[42]}\>{R. T. Sharp, J. Math. Phys., {\bf 16} (1975) 2050}\\
\>{[43]}\>{ A. J. Macfarlane, L. O. O'raifertaigh, and P. S. Rao, J. Math. Phys., {\bf 8} (1967) 536}\\
\>{[44]}\>{G. H. Gadiyar and H. S. Sharatchandra, J. Phys. A, {\bf 25} (1992) L85}\\
\>{[45]}\>{C. K. Chew and R. T. Sharp, Can. J. Phys., {\bf 44} (1966) 2789}\\
\>{[46]}\>{P. Jasselette, J. Phys. A, {\bf 19} (1986) 2261}\\
\>{[47]}\>{M. F. O'Reilly, J. Math. Phys., {\bf 23} (1982) 2022}\\
\>{[48]}\>{B. Preziosi, A. Simoni, and B. Vitale, Nuovo, Cimento, {\bf 34} (1964) 110}\\
\end{tabbing}
\end{document}